\documentclass[conference]{IEEEtran}
\IEEEoverridecommandlockouts

\usepackage{cite}
\usepackage{amsmath,amssymb,amsfonts}
\usepackage{graphicx}
\usepackage{textcomp}
\usepackage{xcolor}
\usepackage{multirow}
\usepackage{algorithm}
\usepackage{algpseudocode}
\usepackage{url}
\usepackage{graphics}
\usepackage{xcolor}
\usepackage{microtype}
\begin{document}
\newcommand{\pretrainingshort}{DAS~}
\def\x{{\mathbf x}}
\def\L{{\cal L}}

\newcommand{\blue}[1]{\textcolor{blue}{#1}}
\newcommand{\red}[1]{\textcolor{red}{#1}}
\newcommand{\relgreen}[1]{\textcolor{teal}{\small(#1)}}
\newcommand{\relred}[1]{\textcolor{red}{\small(#1)}}
\title{A Domain Adaptation Framework for Speech Recognition Systems with Only Synthetic data\\
\thanks{This work was done during an internship at Meta.}
}
\author{Minh Tran$^{1*}$, Yutong Pang$^2$, Debjyoti Paul$^2$, Laxmi Pandey$^2$, Kevin Jiang$^2$, Jinxi Guo$^2$, \\Ke Li$^2$, Shun Zhang$^2$, Xuedong Zhang$^2$, Xin Lei$^2$ \\ $^1$ University of Southern California \; $^2$ Meta AI}

\maketitle

\begin{abstract}
We introduce \pretrainingshort(Domain Adaptation with Synthetic data), a novel domain adaptation framework for pre-trained ASR model, designed to efficiently adapt to various language-defined domains without requiring any real data. 
In particular, \pretrainingshort first prompts large language models (LLMs) to generate domain-specific texts before converting these texts to speech via text-to-speech technology. The synthetic data is used to fine-tune Whisper with Low-Rank Adapters (LoRAs) for targeted domains such as music, weather, and sports.
We introduce a novel one-pass decoding strategy that merges predictions from multiple LoRA adapters efficiently during the auto-regressive text generation process. 
Experimental results show significant improvements, reducing the Word Error Rate (WER) by $10\%$ to $17\%$ across all target domains compared to the original model, with minimal performance regression in out-of-domain settings $-1\%$ on Librispeech test sets.
We also demonstrate that \pretrainingshort operates efficiently during inference, introducing an additional $9\%$ increase in Real Time Factor (RTF) compared to the original model when inferring with three LoRA adapters.
\end{abstract}

\section{Introduction}
Current state-of-the-art speech recognition systems are trained on extensive and varied datasets. For example, the Whisper model \cite{radford2023robust} underwent training with 680,000 hours of multilingual data, encompassing a wide array of audio conditions and environments. This extensive training strategy has enabled these models to express a high level of robustness in speech recognition across diverse conditions and multiple evaluation datasets \cite{radford2023robust}. Such robustness motivates us to question whether domain-specific data distribution, especially transcript distribution over language characterizing a given topic such as music or sports, are still relevant. Are state-of-the-art speech recognition systems universally effective, or is there still benefits in tailoring these robust models to specific domains?

In this paper, we explore this research question using Whisper, one of the most widely utilized and state-of-the-art automatic speech recognition (ASR) models. In particular, we introduce a domain adaptation framework, which we refer to as \pretrainingshort. The framework is designed to efficiently adapt Whisper \cite{radford2023robust} to various language-defined domains, \textit{i.e.,} speech with corresponding transcripts characterizing a given topic or genre. \pretrainingshort is characterized by two key features: 1) it does not require any real data (audio or text) for adaptation to a target domain, thereby eliminating the substantial costs of human annotations and enhancing scalability, and 2) it maintains effective adaptation across different domains without compromising its generalizability in out-of-domain conditions. 

To achieve the first goal, we prompt large language models (LLMs) to generate texts that are relevant to a target domain. Subsequently, we employ text-to-speech to convert the generated text into corresponding audio data, and use this data to fine-tune the Whisper model specifically for that domain. To fulfill the second goal, we explore the use of a single encoder with multiple decoders, with each decoder dedicated to a specific domain. To mitigate the additional computational costs associated with multiple decoders, we employ LoRA (Low-Rank Adaptation) adapters. Additionally, we propose a novel one-pass decoding strategy that efficiently merge predictions from multiple LoRA adapters during each step of the auto-regressive text generation process. 
Experimental results demonstrate that our model consistently achieves a reduction ranging from $10\%$ to $17\%$ in the Word Error Rate metric (WER) over the original Whisper model on real speech across three target domains: music, weather/datetime, and sports. Notably, the model exhibit minimal performance regression in out-of-domain settings on multiple public benchmarks while introducing a modest latency increase of $9\%$ in the Real Time Factor (RTF) metric compared to the original model. In summary, our contributions include:
\begin{itemize}
    \item We provide a novel and thorough analysis on the effect of adapting ASR systems to language-defined domains using only synthetic audio/text data.
    \item We propose \pretrainingshort, a novel domain adaptation framework for speech recognition systems, that effectively and efficiently adapt a base model to a set of target domains without sacrificing generalizabilty.
    \item We provide extensive experiments and ablation study to demonstrate the effectiveness of \pretrainingshort.
\end{itemize}
\section{Related work}
\begin{figure*}[t]
\centering
\includegraphics[width=\linewidth, trim={20pt 150pt 0pt 80pt}, clip]{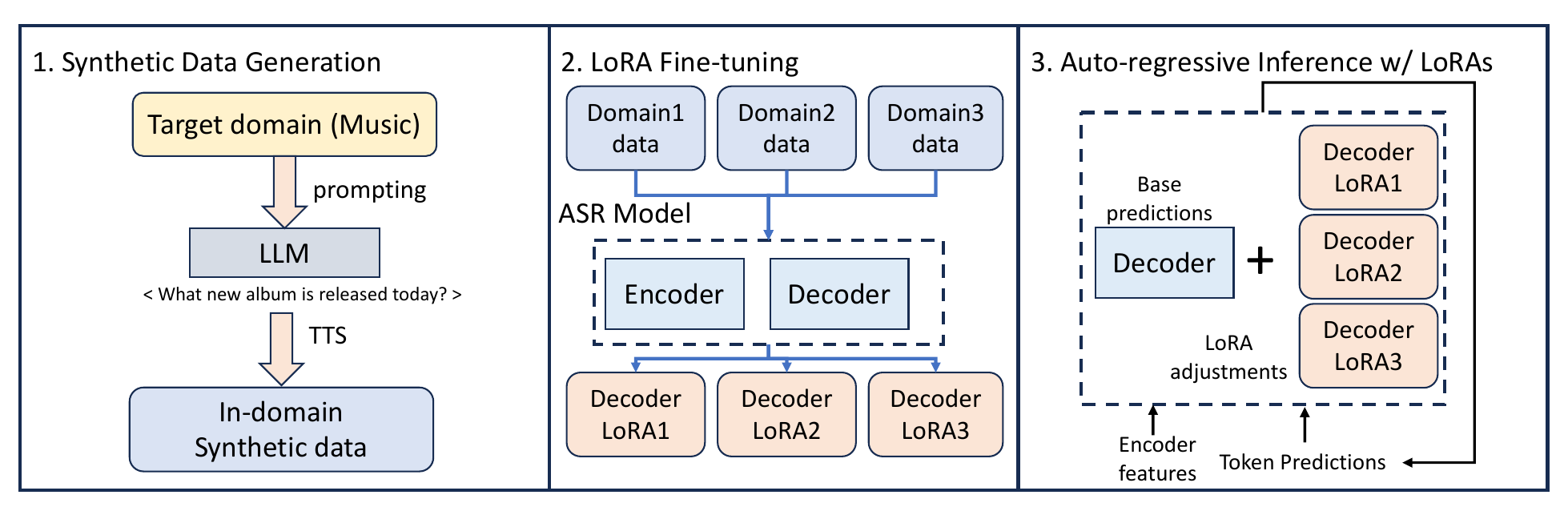}
\vspace{-4mm}
\caption{\textbf{An overview of the proposed domain adaptation framework.} We first use LLM to generate texts related to a particular domain of interest (\textit{e.g.,} music) and then use text-to-speech (TTS) to generate synthetic speech. We then fine-tune the decoder of \textit{Whisper} with the generated synthetic data using Low-rank Adaptation (LoRA) \cite{hulora}. Finally, we propose \pretrainingshort to effectively generate predictions with multiple LoRAs to enhance in-domain speech recognition performance with minimal regression in out-of-domain settings.
}
\label{fig:framework}
\vspace{-10pt}
\end{figure*}
Existing work on domain adaptation for ASR mainly focus on unsupervised domain adaptation settings, where both source and target domain speech are available but word transcripts are only available in the source domain. Sun \textit{et al.} \cite{sun2017unsupervised} proposed using domain adversarial learning with Gradient Reversal Layer \cite{ganin2015unsupervised} on deep neural networks to handle noise robustness or speech with different genres. Hsu \textit{et al.}\cite{hsu2017unsupervised} propose a data augmentation method based on latent features extracted from variational autoencoders learned from both source and target domain speech.  Meng \textit{et al.} \cite{meng2019domain} extends Teacher/Student (T/S) learning to large-scale unsupervised domain adaptation of an end-to-end (E2E) model by using teacher's token posteriors as soft labels and one-best predictions for decoder guidance. Samarakoon \textit{et al.}\cite{samarakoon2018domain} explore domain adaptation for ASR in low-resource setting by adding domain labels to the label sequences, using factorized hidden layer adaptation and a novel domain-specific gating mechanism. Sim \textit{et al.} \cite{sim2018domain} propose using factorized hidden layer for LSTM. Sim \textit{et al.}\cite{sim2024comparison} present an extensive study of different parameter-efficient fine-tuning methods for domain adaptation, and show the superior performance of Residual Adapters \cite{rebuffi2017learning} and Low-rank Adapters (LoRA) \cite{hulora}. 
Zheng \textit{et al.} \cite{zheng2021using} explore the use of synthetic speech to enhance recognition performance on out-of-vocabulary words. In contrast, our work focuses on domain adaptation, aiming to increase the likelihood of specific vocabulary distributions being accurately predicted.
Most relevant to our work, Huang et al. \cite{huang2023text} fine-tune BART on a seed set of in-domain texts to generate domain-relevant texts. They use TTS-generated speech from these texts to augment the LibriSpeech dataset and train an RNN-T based model \cite{le2021contextualized} from scratch, demonstrating significant improvement in in-domain ASR performance. 

Compared to existing work, our main contributions are: 1) \pretrainingshort does not assume any available in-domain text or real audio data, allowing scalability; 2) unlike \cite{huang2023text}, we focus on adapting a pre-trained ASR model and use only synthetic speech for adaptation rather than treating it as augmentation data alongside real speech data; and 3) we are the first to explore strategies to avoid performance regression for out-of-domain speech for ASR systems. 

\section{Method}
\noindent \textbf{Problem Formulation.} Given a pre-trained ASR model $\mathcal{M}$ and a set of language-defined domains $\{d_1, \dots, d_k\}$ with corresponding real speech test sets $\{\mathcal{D}_1, \dots, \mathcal{D}_k\}$, our goal is to produce a new model $\mathcal{M'}$ from $\mathcal{M}$ such that $\mathcal{M'}$ achieves better performance (lower Word-Error-Rate) than $\mathcal{M}$ across all in-domain test sets while maintaining comparable performance to $\mathcal{M}$ on any out-of-domain test sets. In this study, we focus on three real language-defined test sets pertaining to Music, Weather/Date-time, and Sports. Further details of the datasets are provided in Section \ref{sec:datasets}.
\begin{table}[t]
\centering
\begin{tabular}{l|l} 
\hline
\multirow{3}{*}{\rotatebox{90}{Music}}   & \begin{tabular}[c]{@{}l@{}}What musical movement did the beatles define in \\ the sixties\end{tabular}                          \\
                         & \begin{tabular}[c]{@{}l@{}}What group was Stevie Nicks a part of before\\ Fleetwood Mac\end{tabular}                            \\
                         & What's the title of elvis presley's first studio album                                                                          \\ \hline
\multirow{3}{*}{\rotatebox{90}{Weather}} & \begin{tabular}[c]{@{}l@{}}What's the average temperature difference between \\ winter and summer in this park?\end{tabular}    \\
                         & What's the current temperature outside right now?                                                                               \\
                         & \begin{tabular}[c]{@{}l@{}}How many centimeters of snow have fallen in Paris so \\ far this month?\end{tabular}                 \\ \hline
\multirow{3}{*}{\rotatebox{90}{Sports}}  & \begin{tabular}[c]{@{}l@{}}Which boxer held the heavyweight title for the longest \\ duration in the 20th century?\end{tabular} \\
                         & \begin{tabular}[c]{@{}l@{}}Which team did Tom Brady lead to victory in Super \\ Bowl XXXVI?\end{tabular}                        \\
                         & \begin{tabular}[c]{@{}l@{}}Wow many consecutive games did Cal Ripken Jr. play \\ in the MLB?\end{tabular}   \\\hline                   
\end{tabular}
\vspace{1mm}
\caption{Samples of chatbot-style synthetic texts generated from our in-domain text generation pipeline with \texttt{Llama3-70b} \cite{llama3modelcard}.}
\label{tab:samples}
\vspace{-18pt}
\end{table}
\subsection{Synthetic data generation} We use the \texttt{Llama3-70b} model \cite{llama3modelcard} to generate synthetic text for a specific domain, eliminating the need for a seed text dataset as in \cite{huang2023text}. To enhance the diversity of the generated texts, our data generation process consists of two phases. In the first phase, we utilize LLMs to create tailored instructions related to the domain of interest based on some input seed instructions. Our seed instructions include simple fields such as \textit{usecase} (e.g., question generation in English), \textit{skills} (e.g., English singer, English album name), \textit{persona} (e.g., teenager), and \textit{instruction} (e.g., generate chatbot questions related to music). We base our instruction generation pipeline on \textit{CodecLM} \cite{wang2024codeclm}, which initially encodes seed instructions into metadata (keywords generated on-the-fly) and then decodes the metadata to produce tailored instructions. These instructions are then fed into the LLMs to generate in-domain texts. 
Table \ref{tab:samples} show some samples of our text generation pipeline. We then use an internal text-to-speech system to convert the generated text into synthetic speech data. We only use a single target speaker for all synthetic data without any audio augmentation.

\subsection{\pretrainingshort}
\noindent \textbf{Decoder-only fine-tuning.} 
\begin{table}[t]
\centering
\begin{tabular}{c|ccc}
\hline
           & \multicolumn{1}{l}{music} & \multicolumn{1}{l}{weather} & sports                     \\ \hline
original   & 27.94                      & 14.97                        & \multicolumn{1}{r}{15.59} \\ 
ft enc-dec & 25.89                      & 16.72                        & 16.23                          \\ 
ft dec (last 3 layers)    & 24.99                       & 14.45                       & 15.46 \\
ft full dec (6 layers)    & \textbf{23.20}                       & \textbf{12.88}                        & \textbf{15.30}                          \\ \hline
\end{tabular}
\vspace{1mm}
\caption{Comparison (WER metric) of different fine-tuning components using TTS in-domain speech data with the \textit{Whisper-base} model on 3 domain-specific datasets.}
\label{tab:dec_ft}
\vspace{-20pt}
\end{table}
Unlike previous approaches that treat synthetic data as an augmentation to a real speech dataset and train models from scratch \cite{huang2023text}, we face challenges when working with pre-trained ASR models, where it is costly to re-train the models and the pre-training dataset is sometimes unavailable as in the case of Whisper \cite{radford2023robust}. Furthermore, using a specific dataset such as LibriSpeech \cite{panayotov2015librispeech} as the real counterpart might cause the pre-trained model to overfit to the specific conditions of that dataset (\textit{e.g.}, clean audio), potentially compromising its generalizability. To address these issues, we explore different strategies for \textit{Whisper-base} when fine-tuned solely with TTS-generated data. The results, presented in Table \ref{tab:dec_ft}, indicate that fine-tuning with just the decoder proves most advantageous when no real data is accessible. This finding leads us to develop \pretrainingshort to use a single version of the pre-trained encoder followed by multiple versions of the decoder, with each version tailored to a specific domain. More details are provided in Section \ref{sec:implementation}.\\
\noindent \textbf{Parameter-efficient fine-tuning.} The decoder accounts for the majority of the parameters in the Whisper model ($72.4\%$ in the \texttt{Whisper-base} version). Maintaining multiple versions of the decoder for each specific domain could be prohibitively expensive. Hu \textit{et al.} \cite{hulora} show that adapting LLMs can be done in a storage and computation efficient manner by integrating low rank decomposition matrices into the Transformer architecture \cite{vaswani2017attention}. In particular, for a pre-trained weight matrix $W_o$ in a pre-trained neural network, the fine-tuning update $\Delta W$ can be constrained by two light-weighted matrices $A$ and $B$ such that 
\begin{equation}
    h = W_0x + \Delta W x = W_0x + BAx
\end{equation}
During training, the pre-trained weights $W_0$ remain frozen and only $A$ and $B$ are trainable, allowing for efficient learning and adaptation. In this paper, we employ one LoRA adapter with less than $2\%$ of the total number of parameters in Whisper for each domain to adapt the decoder of the Whisper model. 
\\
\noindent \textbf{Auto-regressive decoding with multiple LoRAs.}
\begin{algorithm}[t]
\caption{Auto-regressive decoding with multiple LoRAs}\label{alg:cap}
\small
\begin{algorithmic}[1]
\Require $W$, $\{(A_i, B_i)\}$ for $i \in [k]$, $x$: encoder features
\State $tokens \gets []$
\While{$[eos] \notin tokens$}
\State h = $Softmax(W(x,tokens))$
\State (next$_0$, c$_0$) = Argmax$(h)$, Max$(h)$ \Comment{c denotes the confidence}
\State $h_i$ = Softmax$((W+B_iA_i)(x, tokens))$ for $i \in [k]$ 
\State (next$_i$, c$_i$) = Argmax$(h_i)$, Max$(h_i)$ for $i \in [k]$
\State SELECT next from \{next$_0$, next$_1$,\dots, next$_k$\} 
\State INSERT next to $tokens$
\EndWhile\\
\Return $tokens$

\end{algorithmic}
\label{alg:autoregressive}
\end{algorithm}
We introduce Auto-regressive Decoding with Multiple LoRA (ADML), detailed in Algorithm \ref{alg:autoregressive}. ADML enables a single decoding pass that incorporates multiple LoRA adapters, efficiently leveraging batched computations with the base model weights \( W \), as described in \cite{sheng2023s}. To optimize LoRA computations \( B_iA_i(x, \text{tokens}) \), we restructured them from a sequential process into a batched computation, as follows: 
\begin{equation}
\{B_iA_ix_i\}_{i=1}^N=B(AX)
\end{equation}
where $A=concat(A_1,\dots,A_k)$, $X=concat(x_1,\dots,x_k)$, and $B=blk\_diag(B_1,\dots,B_k)$\footnote{\url{https://pytorch.org/docs/stable/generated/torch.block_diag.html}}. Next, we select the optimal next token based on the confidence levels produced by different LoRA weights in Line 7. Through experiments with various next word selection strategies, we find out that it is best to select the next word from the $k+1$ predictions according to
\begin{equation}
    max\{c_i\}_{i=0}^k-c_0 \geq \tau \;\; OR \;\; min\{c_i\}_{i=0}^k-c_0 \leq -\tau 
\end{equation}
where $\tau$ is a threshold parameter. If both conditions are met, we priortize the word with the maximum confidence. If none of the conditions are met, we pick the word predicted by the original model.\\
\vspace{-10pt}

\section{Experiments}
\noindent \textbf{Datasets} \label{sec:datasets}
\begin{table}[t]
\centering
\begin{tabular}{c|ccc}
\hline
           & \multicolumn{1}{l}{music} & \multicolumn{1}{l}{weather} & sports                     \\ \hline
Synthetic dataset& 44K                      & 31K                        & \multicolumn{1}{r}{46K} \\ 
Evaluation dataset     & 2.1K                       & 2.8K                       & 5.1K                          \\ \hline
\end{tabular}
\vspace{1mm}
\caption{\# of samples for each dataset used in this study.}
\label{tab:stats}
\vspace{-25pt}
\end{table}
\begin{table*}[t]
\centering
\begin{tabular}{ll|lll}
\hline
               & Train set   & \multicolumn{1}{l}{music} & \multicolumn{1}{l}{weather} & \multicolumn{1}{l}{sports} \\ \hline
Original       & -           & 27.94                      & 14.97                        & 15.59                       \\ 
FT             & TTS-Music   & \textbf{23.20} \relgreen{$\uparrow17.0\%$}                      & 14.45 \relgreen{$\uparrow3.5\%$}                        & 20.1 \relred{$\downarrow28.8\%$}                      \\ 
FT             & TTS-Weather & 33.05 \relred{$\downarrow18.3\%$}                     & \textbf{12.10} \relgreen{$\uparrow19.2\%$}                         & 17.7 \relred{$\downarrow13.5\%$}                         \\ 
FT             & TTS-Sports  & 25.05 \relred{$\downarrow10.3\%$}                      & 15.96 \relred{$\downarrow6.6\%$}                        & \textbf{15.3} \relgreen{$\uparrow1.9\%$}                        \\ \hline
LoRA-ft        & TTS-Music   & \textbf{23.23} \relgreen{$\uparrow16.8\%$}                      & 13.27 \relgreen{$\uparrow11.3\%$}                         & 16.51 \relred{$\downarrow5.9\%$}     \\ 
LoRA-ft        & TTS-Weather & 26.65 \relgreen{$\uparrow4.6\%$}      & \textbf{11.70} \relgreen{$\uparrow21.8\%$}                         & 15.08 \relgreen{$\uparrow 3.3\%$}    \\ 
LoRA-ft        & TTS-Sports  & 27.14 \relgreen{$\uparrow2.9\%$}      & 14.05 \relgreen{$\uparrow6.1\%$}      & \textbf{13.37}  \relgreen{$\uparrow14.2\%$}                     \\ \hline
FT-Multi       & TTS(M+W+S) & \textbf{24.71} \relgreen{$\uparrow11.6\%$}                     & 24.53 \relred{$\downarrow64.0\%$}                       & 15.84   \relred{$\downarrow1.6\%$}                    \\ 
LoRA-ft-Multi  & TTS(M+W+S) & 25.09 \relgreen{$\uparrow10.2\%$}    & 13.70 \relgreen{$\uparrow8.4\%$}       & 14.61 \relgreen{$\uparrow6.3\%$}     \\ 
\pretrainingshort           & TTS(M/W/S) & 24.87 \relgreen{$\uparrow11.0\%$}                     & \textbf{12.39} \relgreen{$\uparrow17.2\%$}                       & \textbf{13.98}  \relgreen{$\uparrow10.3\%$}                     \\ \hline
\end{tabular}%
\vspace{1mm}
\caption{ASR performance comparison between \pretrainingshort and baselines across three real domain-specific test sets.}
\label{tab:main_results}
\vspace{-20pt}
\end{table*}
For evaluation, we utilize three real speech internal datasets collected using the Meta Ray Ban Glasses, with each sample being manually annotated. These datasets are divided into three language-defined domains: Music, Weather, and Sports. Given the nature of the data, the content predominantly features chatbot-style interactions, where a user requests Meta AI to perform tasks that involve verifying information via the internet. Statistics of the generated synthetic (for training) and evaluation datasets are available in Table \ref{tab:stats}. \\
\noindent \textbf{Implementation details} \label{sec:implementation}
We use \texttt{Whisper-base} as the pre-trained ASR model in this study. \texttt{Whisper-base} architecture resembles a vanilla transformer \cite{vaswani2017attention}, with $6$ layers of transformer encoder, $6$ layers of transformer decoder and a feature dimension of $512$ for a total of $74M$ parameters. For LoRA, we use the bottleneck dimension $r=32$, $\alpha=64$ with Rank-stable scaling \cite{kalajdzievski2023rank}, PiSSA initialization \cite{meng2024pissa}, and add LoRA adapters to the query and value attention blocks of Whisper, which are determined via hyper-parameter tuning. This results in around $1.2\%$ additional parameters for each adapter added to the base model. During training, we fine-tune each LoRA weights independently with an in-domain synthetic dataset. The LoRA adapters are trained with an AdamW optimizer with $lr=3e^{-6}$ for $10$ epochs with a batch size of $16$. We use a linear warmup learning rate for $10\%$ of the training process. For inference with multiple LoRAs, we use $\tau=0.025$. For evaluation, we report the Word Error Rate (WER) metric without punctuations and wakewords. \\
\noindent \textbf{Baselines} We compare our model with the following baselines: \textbf{FT} - (Decoder) vanilla finetuning; \textbf{LoRA-ft} - (Decoder) finetuning with LoRA adapters; \textbf{FT-Multi} - Full finetuning Whisper's decoder parameters on all 3 domains by combining the generated TTS data, where each training batch is drawn equally from the 3 domains, and \textbf{\pretrainingshort} - Our proposed framework.

\vspace{-2mm}
\section{Results}
\begin{table}[t]
\centering
\begin{tabular}{l|cccc}
\hline
                      & \multicolumn{1}{l}{$OOD_1$} & \multicolumn{1}{l}{$OOD_2$} & \multicolumn{1}{l}{$OOD_3$} & \multicolumn{1}{l}{$OOD_4$} \\ \hline
Original              & 12.02                                       & 5.04             & 10.87    & 10.36                                                   \\ 
LoRA-ft Multi                  & 13.79                                        & 5.78   & 11.48       & 10.82                                                       \\ 
\
\pretrainingshort                  & 12.25                                        & 5.1   & 11.06       & 10.29                                                       \\ 
\% change & -1.02                                     & -1.01                    & -1.02  & +0.99                                       \\ \hline
\end{tabular}
\vspace{1mm}
\caption{ASR performance comparison between \pretrainingshort and original (unadapted) model across four out-of-domain test sets. $OOD_1$: LibriSpeech test-other, $OOD_2$: LibriSpeech test-clean, $OOD_3$: Fleurs-En, $OOD_4$: Voxpopuli-En.}
\label{tab:regression}
\vspace{-25pt}
\end{table}
Our evaluation results, presented with the mentioned baselines and \pretrainingshort in Table \ref{tab:main_results}, demonstrate that fine-tuning the decoder with LoRAs \cite{hulora} can potentially surpass the performance of full-finetuning (FT). Notably, LoRA-ft in the Music domain shows comparable results to (FT) with a WER of around $23.2\%$ while outperforming FT in the remaining two domains. This aligns with findings reported in \cite{hulora}. 

It is also important to highlight that both FT and LoRA-FT exhibit performance regression in domains other than the one they were adapted to. For example, FT adapted to the music domain (TTS-M) achieves a WER of $23.3\%$, which is a substantial improvement of $16.9\%$ for the Music domain. However, it also results in a significant regression in the Sport domain, with a relative WER increase of $28.9\%$ compared to the original model. 

\pretrainingshort addresses this issue with the ADML component, which efficiently encodes transcripts in a single decoding pass using multiple domain-specific adapters, leading to consistent improvements of $11.0\%$, $17.2\%$, and $10.3\%$ across the music, weather, and sports domains, respectively. Further comparisons of \pretrainingshort with FT-Multi and LoRA-ft-Multi—two baselines fine-tuned on a combination of all three TTS datasets—demonstrate that \pretrainingshort still maintains superior performance. We further evaluate \pretrainingshort in a completely out-of-domain setting using the \textit{LibriSpeech test-clean} and \textit{test-other} test sets, which feature vocabulary unrelated to that chatbot language that \pretrainingshort was adapted to. As demonstrated in Table \ref{tab:regression}, \pretrainingshort maintains performance comparable to the original Whisper model, with only about a $1\%$ regression in performance. This demonstrates its robustness and resistance to out-of-domain regression.

\begin{table}[t]
\centering
\begin{tabular}{l|cccc}
\hline
               & \multicolumn{1}{c}{RTF} & \multicolumn{1}{c}{$\% \Delta_P$} & \multicolumn{1}{c}{$\% \Delta_S$} & \multicolumn{1}{c}{Speedup}\\ \hline
original  & 0.0244                          & \multicolumn{1}{l}{-} & \multicolumn{1}{l}{-}            & \multicolumn{1}{l}{-}                  \\ 
1-domain       & 0.0269                          & 0.102 & 0.102  & 1x                                  \\ 
2-domain       & 0.0263                         & 0.077   & 0.213   & 2.76x                             \\ 
3-domain       & 0.0267                          & 0.094         & 0.381  & 4.05x                         \\ \hline
10-domain      & 0.0289                          & 0.184    & 1.582  & 8.6x                                  \\ 
25-domain      & 0.0451                          & 0.848      & 3.862        & 4.6x                      \\ \hline
\end{tabular}
\vspace{1mm}
\caption{Latency comparison between \pretrainingshort and original model with different numbers of adapters (domains). We also provide sequential LoRA computations ($\% \Delta_S$) to demonstrate the benefits of parallelization ($\% \Delta_P$) as described in Eq (2).}
\label{tab:latency}
\vspace{-25pt}
\end{table}
With the additional LoRA adapters, \pretrainingshort introduces some extra computations with respect to the orginal auto-regressive decoding process. We benchmark the additional inference latency in Table \ref{tab:latency}. In particular, we run inference on a V100-32GB GPU on $3K$ randomly sampled utterances drawn from the three evaluation datasets and measure the Real-time-factor (RTF), which is the ratio between the processing time and the audio length. 
We also evaluate the scalability of \pretrainingshort for more than three domains by replicating the initial three domains until reaching the desired number of domains. \pretrainingshort introduces a modest inference latency increase of approximately $9\%$ compared to the original model and demonstrates strong scalability as the number of inference domains grows. However, the computational demands rise with additional domains (e.g., 25 domains), potentially resulting in diminishing performance gains on less powerful hardware. We anticipate that leveraging more advanced GPUs, such as A100 or H100, could alleviate these limitations.
\vspace{-5pt}
\section{Conclusion}
In conclusion, we introduce \pretrainingshort, a novel domain adaptation framework for the pretrained ASR systems with only synthetic data. By employing a single encoder with multiple domain-specific decoders finetuned with Low-Rank Adapters (LoRAs), the framework enables better domain-specific performance while maintaining high efficiency and robustness to out-of-domain regression.

\bibliographystyle{IEEEtran}
\bibliography{IEEEabrv,refs}

\end{document}